\documentclass[10pt]{article}

\usepackage{graphicx,epsfig,pstricks,rotating}
\usepackage{graphics}
\usepackage{amssymb}
\usepackage{amsfonts}
\usepackage{amsmath}

\begin{document}
\begin{center}
{\bf DECAY OF A SCALAR $\sigma$-MESON NEAR THE CRITICAL END-POINT  IN THE PNJL MODEL}
\end{center}

\begin{center}
{\it A.V. Friesen, Yu.L. Kalinovsky and V.D. Toneev}\\
Join Institute for Nuclear Research, Dubna
\end{center}

{\small Properties of a scalar $\sigma$-meson are investigated in the
two-flavor Nambu-Jona-Lasinio model with the Polyakov loop. Model
analysis of the phase diagram of strong interacting matter is
performed. The temperature dependence of the $\sigma\rightarrow\pi\pi$
decay width is studied at the zero chemical potential and near the
critical end-point. The calculated strong coupling constant
$g_{\sigma\pi\pi}$ and the decay width are compared with available
experimental data and other model results. Nonthermal enhancement
of the total decay width  is noted for the $\sigma$ meson near the
critical end-point when the condition $m_\sigma\geq2m_\pi$ is
broken.
   }\\

PACS 13.25.Jx, 25.75.Nq

\section*{Introduction}

The models of Nambu--Jona-Lasinio  type \cite{volk2,volk3,Ebert,VolkM}
have a long history
and are used to describe the dynamics and thermodynamics of light mesons.
This type of models gives a simple and practical example  of the basic mechanism
of spontaneous breaking of chiral symmetry and key features of QCD at
finite temperature and chemical potential \cite{echaya,njl2,klev,hatsuda,Eb-kalin}.
The behavior of a QCD system  is governed by the symmetry properties of the
Lagrangian, namely, the global $SU_L(N_f)\times SU_R(N_f)$ symmetry which is
spontaneously broken to $SU_V(N_f)$ and the exact $SU_c(N_c)$ local color symmetry.
On the other hand, in a non-Abelian pure gauge theory, the Polyakov loop
serves as an order parameter of a transition from the low temperature
confined phase ($Z_{N_c}$ symmetric)
to the high temperature deconfined phase characterized by
the spontaneously breaken $Z_{N_c}$ symmetry (PNJL model).
In the PNJL model, quarks are coupled simultaneously to
the chiral condensate and to the Polyakov loop, and the model
includes the features of both the chiral and $Z_{N_c}$ symmetry breaking.
The model reproduces rather successfully lattice data on QCD thermodynamics.
The use of the PNJL model is therefore reasonable for investigating the in-medium
properties of  mesons and their decays \cite{pedro,mesons}.

The aim of this work is the investigation of the meson properties and $\sigma$ 
decay near the critical end-point (CEP). 
In this letter, we discuss the decay process $\sigma \to
\pi\pi$ at the finite temperature $T$ and chemical potential $\mu$ in
the framework the Nambu-Jona-Lasinio  model with the Polykov-loop (PNJL) that is
believed to describe well the chiral properties and simulates a deconfinement
transition.
%{\bf The sigma decay and hadronisation process at zero density in NJL model
%were firstly studied in Ref.~\cite{hadron}}.
Our motivation  here is to elaborate these features in a large region of
the temperature $T$ and quark chemical potential $\mu$,
where  a non-thermal enhancement of pions
due to the $\sigma\to\pi\pi$ decay may take place.

\section{The model and the phase diagram}

We use the two-flavor PNJL model with the following
Lagrangian~\cite{pnjl1,pnjl2,pnjl3}
\begin{eqnarray}\label{pnjl}
\mathcal{L}_{\it PNJL}=\bar{q}\left(i\gamma_{\mu}D^{\mu}-\hat{m}_0
\right) q+ G  \left[\left(\bar{q}q\right)^2+\left(\bar{q}i\gamma_5
\mathbf {\tau} q \right)^2\right]
-\mathcal{U}\left(\Phi[A],\bar\Phi[A];T\right),
\end{eqnarray}
where the covariant gauge derivative $D_\mu \equiv \partial_\mu -iA_\mu$
with $A^\mu = \delta_0^\mu A^0$, $A^0 = -iA_4$ (the Polyakov calibration).
The strong coupling constant is absorbed in the definition of $A_{\mu}$.
At the zero temperature the Polyakov loop field $\Phi$ and the quark field
are decoupled. Here, the quark field   $\bar{q} = (\bar{u},\bar{d})$, current
masses $\hat{m} =\mbox{diag} (m_u, m_d)$, Pauli matrices 
$\mathbf {\tau}=\sigma/2$ act in the two color flavor space and $G$  is the 
coupling constant.

The gauge sector of the Lagrangian density (\ref{pnjl}) is described by an
effective potential $\mathcal{U}\left(\Phi[A],\bar\Phi[A];T\right)\equiv
\mathcal{U}\left(\Phi,\bar\Phi;T\right)$
\begin{eqnarray}\label{effpot}
\frac{\mathcal{U}\left(\Phi,\bar\Phi;T\right)}{T^4}
&=&-\frac{b_2\left(T\right)}{2}\bar\Phi \Phi-
\frac{b_3}{6}\left(\Phi^3+ {\bar\Phi}^3\right)+
\frac{b_4}{4}\left(\bar\Phi \Phi\right)^2~,
\end{eqnarray}
where
\begin{eqnarray}
\label{Ueff}
b_2\left(T\right)&=&a_0+a_1\left(\frac{T_0}{T}\right)+a_2\left(\frac{T_0}{T}
\right)^2+a_3\left(\frac{T_0}{T}\right)^3~.
\end{eqnarray}
The parameter set is obtained by fitting the lattice results in the pure $SU(3)$
gauge theory at $T_0 = 0.27$ GeV \cite{pnjl2,pnjl3}  and is given in Table
\ref{table1}.

%\begin{center}
\begin{table}[h]
\begin{center}
\begin{tabular}{|c|c|c|c|c|c|}
\hline
$a_0$ & $a_1$ & $a_2$ & $a_3$ & $b_3$ & $b_4$ \\
\hline
6.75  & -1.95 & 2.625 & -7.44 & 0.75  &   7.5 \\
\hline
\end{tabular}
\caption{The parameter set of the effective potential
$\mathcal{U}(\Phi, \overline{\Phi}; T)$.
} \label{table1}
\end{center}
\end{table}
%\end{center}

Before discussing the meson properties, one should indroduce 
the gap equation for constituent quark mass should be introduced.
For describing  the system properties at the finite temperature and density
the grand canonical potential in the Hartree approximation is considered
\cite{pnjl2,pnjl3}
 \begin{eqnarray} \label{grandcan}
\Omega (\Phi, \bar{\Phi}, m, T, \mu) &=&
\mathcal{U}\left(\Phi,\bar\Phi;T\right) + N_f \frac{(m-m_0)^2}{4G}-
2N_c N_f \int_\Lambda \dfrac{d^3p}{(2\pi)^3}
E_p \nonumber \\
&& - 2N_f T \int \dfrac{d^3p}{(2\pi)^3} \left[ \ln N_\Phi^+(E_p)+
\ln N_\Phi^-(E_p) \right]~,
\end{eqnarray}
where $E_p$ is the quark energy, $E_p=\sqrt{{\bf p}^2+m^2}$, $E_p^\pm
= E_p\mp \mu$, and
\begin{eqnarray}
&& N_\Phi^+(E_p) = \left[ 1+3\left( \Phi +\bar{\Phi} e^{-\beta
E_p^+}\right) e^{-\beta E_p^+} + e^{-3\beta E_p^+}
\right]^{-1},  \\
&& N_\Phi^-(E_p) = \left[ 1+3\left( \bar{\Phi} + {\Phi} e^{-\beta
E_p^-}\right) e^{-\beta E_p^-} + e^{-3\beta E_p^-} \right]^{-1}.
\end{eqnarray}
Integrals in Eq. (\ref{grandcan}) contain the three-momentum cutoff $\Lambda$.

From the grand canonical potential $\Omega$ the equations of motion can be obtained
\begin{equation} \label{set}
 \dfrac{\partial \Omega}{\partial m} = 0, \,\,
 \dfrac{\partial \Omega}{\partial \Phi} = 0, \,\,
 \dfrac{\partial \Omega}{\partial \bar{\Phi}} = 0,
\end{equation}
 and the gap equation for the constituent quark mass can be written as follows:
\begin{eqnarray} \label{gap-eq}
m = m_0-N_f G <{\bar q}q> =m_0 + 8 G N_c N_f \int_{\Lambda} \dfrac{d^3p}{(2\pi)^3}
\dfrac{m}{E_p} \left[ 1 - f^+_{\Phi} - f^- _{\Phi}\right]~,
\end{eqnarray}
 where $f_\Phi^+$, $f_\Phi^-$ are the modified Fermi functions
\begin{eqnarray}
f_\Phi^+ = ((\Phi + 2\bar{\Phi}e^{-\beta E^+})e^{-\beta E^+}
+e^{-3\beta E^+})N_\Phi^+, \nonumber\\
f_\Phi^- = ((\bar{\Phi} + 2{\Phi}e^{-\beta E^-})e^{-\beta E^-}
+e^{-3\beta E^-})N_\Phi^-,
\label{fermi}
\end{eqnarray}
 with $E^\pm = E\mp\mu$.
The regularization parameter $\Lambda$,  the  quark current mass $m_0$, the coupling
strength G and physics quantities to fix these parameters are presented in
Table~\ref{param}.

\begin{table}[h]
\begin{center}
\begin{tabular}{|c|c|c||c|c|}
\hline
$m_0$ [MeV] & $\Lambda$ [GeV] & $G$ [GeV]$^{-2}$ & $F_\pi$ [GeV] & $m_\pi$  [GeV]  \\
\hline
5.5 & 0.639 & 5.227 &  0.092 & 0.139 \\
\hline
\end{tabular}
\caption{The model parameters and quantities used for their tuning.}
 \label{param}
 \end{center}
 \end{table}
 The $\sigma$ and $\pi$ meson masses are the solutions of the equation
\begin{equation}
1 - 2G \ \Pi_{{ps}/{s}}(k^2) = 0,
\end{equation}
where $k^2 = m^2_\pi$ and $k^2 = m^2_\sigma$ in pseudo scalar and scalar sectors,
and $\Pi_{{ps}/{s}}$ are standard mesonic correlation functions \cite{klev}
\begin{eqnarray}
&& i\Pi_\pi (k^2) = \int \frac{d^4p}{(2\pi)^4} \ \mbox{Tr}\,
\left[ i \gamma_5 \tau^a S(p+k) i \gamma_5 \tau^b S(p)
 \right],  \label{Polpi} \\
 && i\Pi_\sigma  (k^2)= \int \frac{d^4p}{(2\pi)^4} \ \mbox{Tr}\, \left[
 i S(p+k)i S(p)
 \right].
  \label{Polsig}
\end{eqnarray}
 Both pion-quark $g_{\sigma\pi\pi}(T,\mu)$ and sigma-quark $g_{\sigma\pi\pi}$
coupling strengths can be obtained from $\Pi_{{ps}/{s}}$:
\begin{eqnarray}
g_{\pi /\sigma}^{-2}(T, \mu) = \frac{\partial\Pi_{{\pi}/{\sigma}}(k^2)}
{\partial k^2}\vert_{^{k^2 = m_\pi^2}_{k^2 =m_\sigma^2} }.
\label{couple}
\end{eqnarray}

\begin{figure}[t]
\centerline{
\psfig{file = 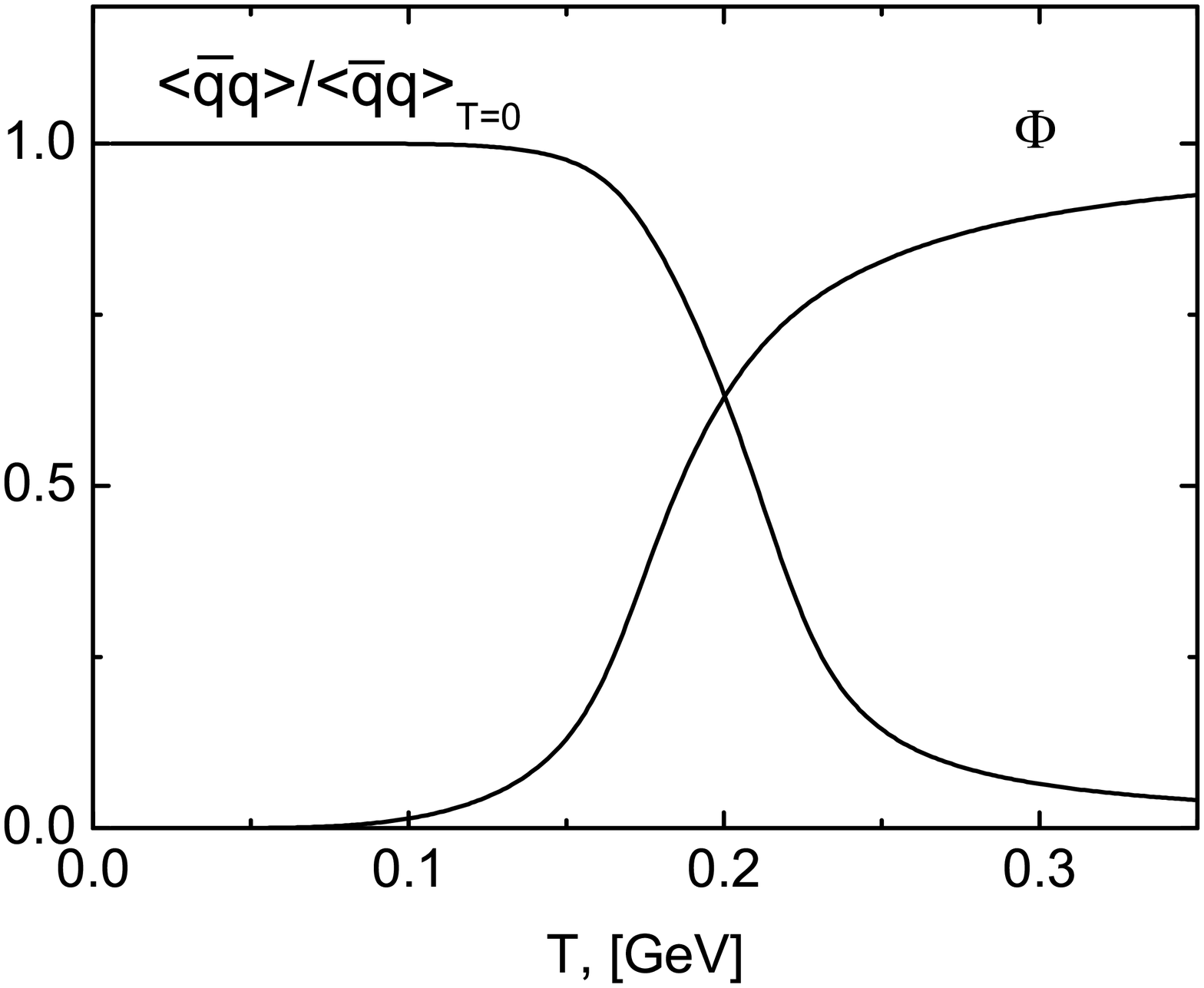, height = 5.cm}\hspace*{-1.0cm}
\psfig{file = 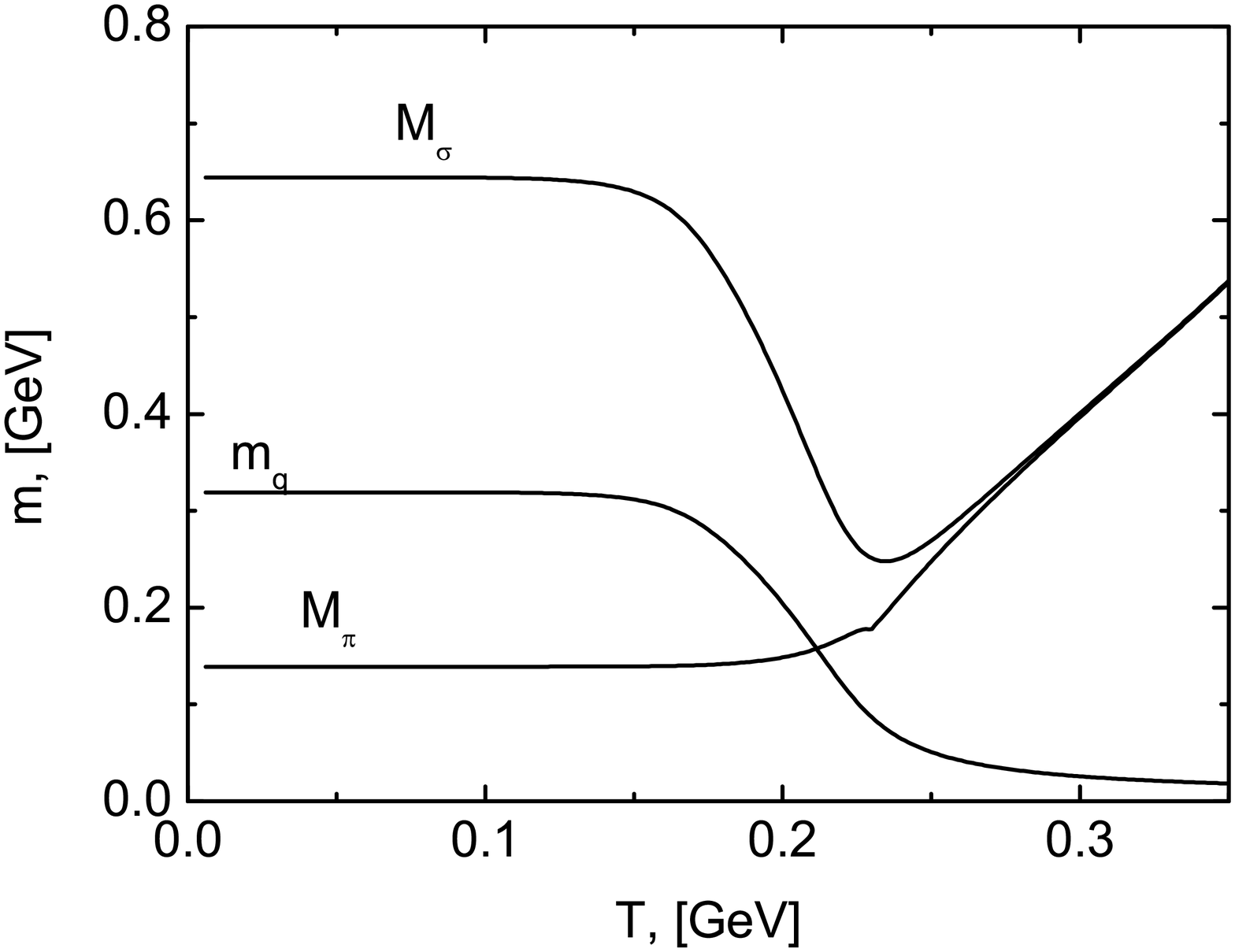, height = 5.0 cm}
}
\caption{ The temperature dependence of the chiral condensate and Polyakov loop
 (left panel) and particle
masses (right panel) at $\mu$ = 0  within the PNJL model. }
 \label{cond}
\end{figure}

As is seen from the gap equation (\ref{gap-eq}), the quark condensate
$<\bar{q}q>$ defines
completely  the quark mass in a hot and dense matter. This correlation
is clearly demonstrated in Fig.~\ref{cond}, where the temperature dependence of
the order parameters of the  chiral condensate and the Polyakov loop, as well
as the sigma and pion masses are shown for  $\mu=0$. The pion hardly changes
its mass and starts to become heavier only near $T_c \sim$ 200 MeV, while the sigma
mass $m_\sigma(T)$  decreases as the chiral symmetry gets restored,
and eventually
 $m_\pi(T)$ becomes larger than double quark mass $m_q$ at the temperature
 $T_{Mott} \sim$ 190 MeV.

\begin{figure}[h!]
\centerline{
\includegraphics[width = 7.5 cm ]{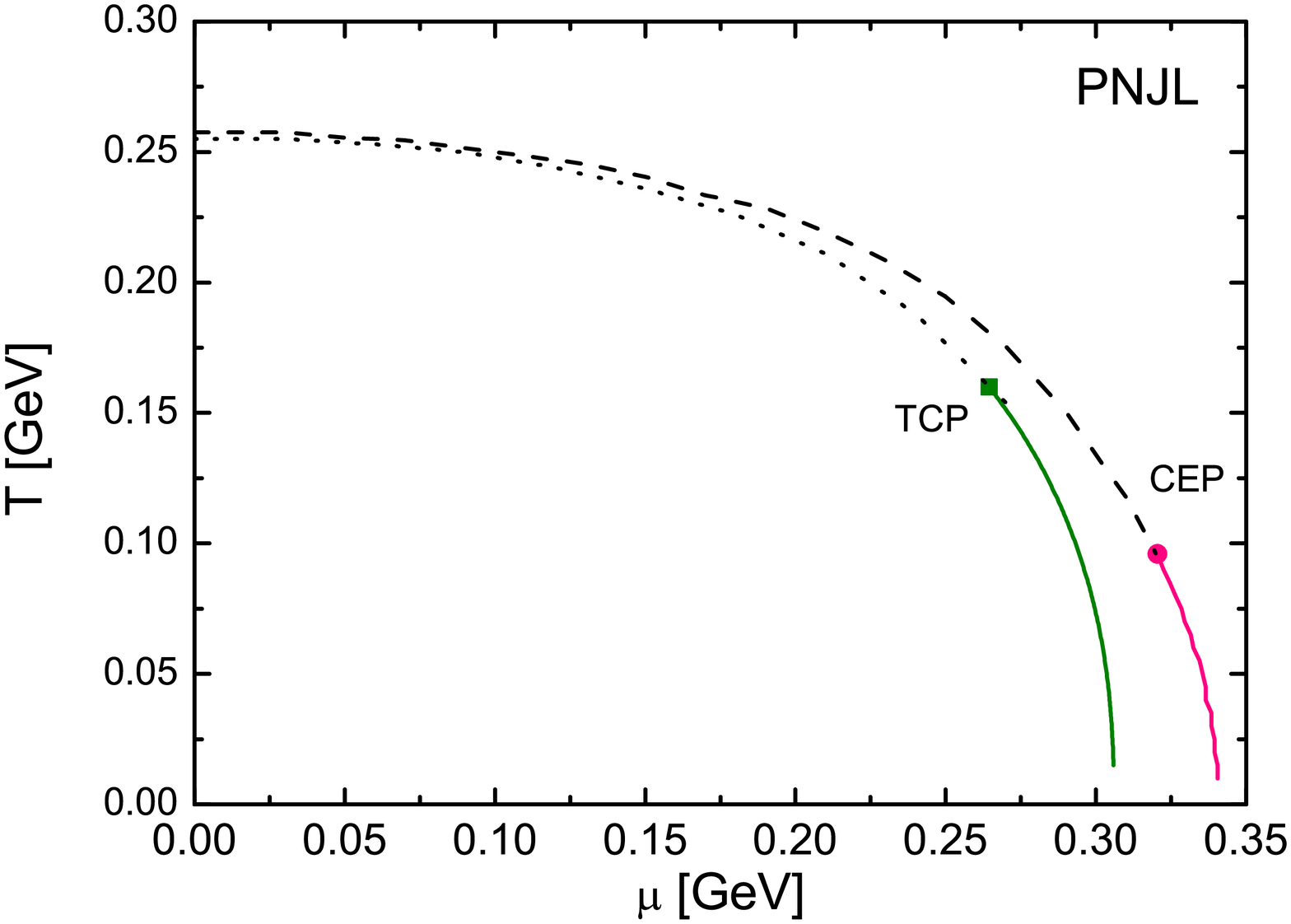}}
\caption{The phase diagram within the PNJL model.
The solid lines denote the first-order transition boundary,
the dotted line is the second-order transition and the dashed line
is a crossover. The lines with the CEP and TCP points are calculated for
 a finite mass and in the chiral limit $m_q=$0, respectively.}
\label{phasediag}
\end{figure}

Since it is still difficult to extract
certain information from the lattice simulations with the nonzero baryon density,
we need QCD models for investigating the phase transitions at the finite baryon 
density. The calculated phase diagram of the physical states of matter within 
the PNJL model is given in Fig.~\ref{phasediag}.
As is seen, in the real world with nonzero pion mass we have  the first-order
phase transition at a moderate temperature and a large baryon chemical potential 
$\mu_B=3\mu$ that, with increasing $T$, terminates at the critical end-point
$(T_{CEP},\mu_{CEP})$ where the second-order phase transition occurs. At a higher
temperature $T>T_{CTP}$ we have a smooth crossover. In the chiral limit 
with massless pions there is a tricritical point that separates
the second-order phase transition at high temperature $T$ and the first-order
transition at lower $T$ and high $\mu$.
For the model parameters  chosen (see Table \ref{param}) we obtain
$T_{CEP} \simeq $0.095 GeV and  $\mu_{CEP} \simeq $0.32 GeV
(cf. with~\cite{FKT11}, where the critical temperature and chemical potential
 are calculated with the same parameters).

\section{Decay $\sigma\rightarrow\pi\pi$}

It is in order here to mention the significance of the scalar meson $\sigma$ (a
chiral partner of the pion) in QCD. A model-independent consequence of
dynamic breaking of chiral symmetry is the existence
of the pion and its chiral partner $\sigma$-meson: The former is the phase
fluctuation of the order parameter ${\bar q}q$ while the latter is the
amplitude fluctuation of ${\bar q}q$.
During the expansion of the system, the in-medium $\sigma$ mass
increases toward its vacuum value and eventually exceeds the $2m_\pi$
threshold. As the $\sigma\to\pi\pi$
coupling is large, the decay proceeds rapidly. Since this process occurs after
freeze-out, the pions generated by it do not have a chance to thermalize. Thus, one
may expect that the resulting pion spectrum should have a nonthermal enhancement
at low transverse momentum.

\begin{figure}[h]
\centerline{
\includegraphics[scale=0.6 ]{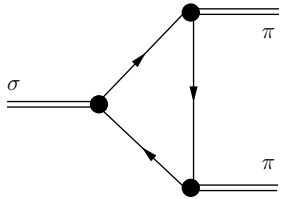}}
\caption{The Feynman diagram of the $\sigma\rightarrow\pi\pi$   decay}.
\label{fdiag}
\end{figure}

To the lowest order in a $1/N_c$ expansion, the diagram for the process
$\sigma\rightarrow\pi\pi$ is shown in Fig.~\ref{fdiag}.
The amplitude of the triangle vertex $\sigma\rightarrow\pi\pi$ can be
obtained analytically as
\begin{eqnarray}
A_{\sigma\pi\pi} = \int \frac{d^4q}{(2\pi)^4} \
Tr \{S(q) \ \Gamma_\pi \ S(q+P) \ \Gamma_\pi \ S(q)\},
\end{eqnarray}
 where $\Gamma_\pi = i\gamma_2\tau$ is the pion vertex function and $S(q) =
{\hat{q} + m}/{\hat{q}^2 - m^2}$ is the quark propagator, a trace is beeng taken
over color, flavor and spinor indices. After tracing and evaluation of
the Matsubara sum one obtains \cite{hadron,zhuang}
\begin{eqnarray}
A_{\sigma\pi\pi} &=& 2mN_cN_f\int\frac{d^3q}{(2\pi)^3}\frac{(1 -
f^+_{\Phi} - f^-_{\Phi})}{2E_q}  \nonumber \\
&\times& \frac{({\bf q}\cdot {\bf p})^2 - (2m_{\sigma}^2 +
4m_{\pi}^2)  ({\bf q}\cdot {\bf p})  + m_{\sigma}^2/2 -
2m_{\sigma}^2E_q^2}{(m_{\sigma}^2 - 4E_q^2)((m_\pi^2 - 2
({\bf q}\cdot {\bf p}))^2 - m_{\sigma}^2E_q^2)}~,
\end{eqnarray}
 where $f_\Phi^+$, $f_\Phi^-$ are the modified Fermi functions (\ref{fermi}).
The coupling strength $g_{\sigma\pi\pi}(T, \mu) = 2g_\sigma
 g^2_\pi A_{\sigma\pi\pi}(T, \mu)$, where $g_\sigma$ and $g_\pi$ are coupling
 constants defined from Eq. (\ref{couple})

The decay width is defined by the cut of the Feynman diagram in
Fig.\ref{fdiag} treating the sigma meson as a quark-antiquark
system
\begin{eqnarray} \label{Gam0}
\Gamma_{\sigma\rightarrow\pi\pi} = \frac{3}{2} \ \frac{g^2_{\sigma\pi\pi}} {16\pi \
m_{\sigma}}
\sqrt{1 - \frac{4m_{\pi}^2}{m_{\sigma}^2}}~.
\end{eqnarray}
The scalar meson $\sigma$  can decay  either into two neutral or two
charged pions. All these channels are taken into account. The factor 3/2 in 
Eq.(\ref{Gam0}) takes into account the isospin conservation.
\begin{figure}
\centerline{
\includegraphics[width = 7.5 cm ]{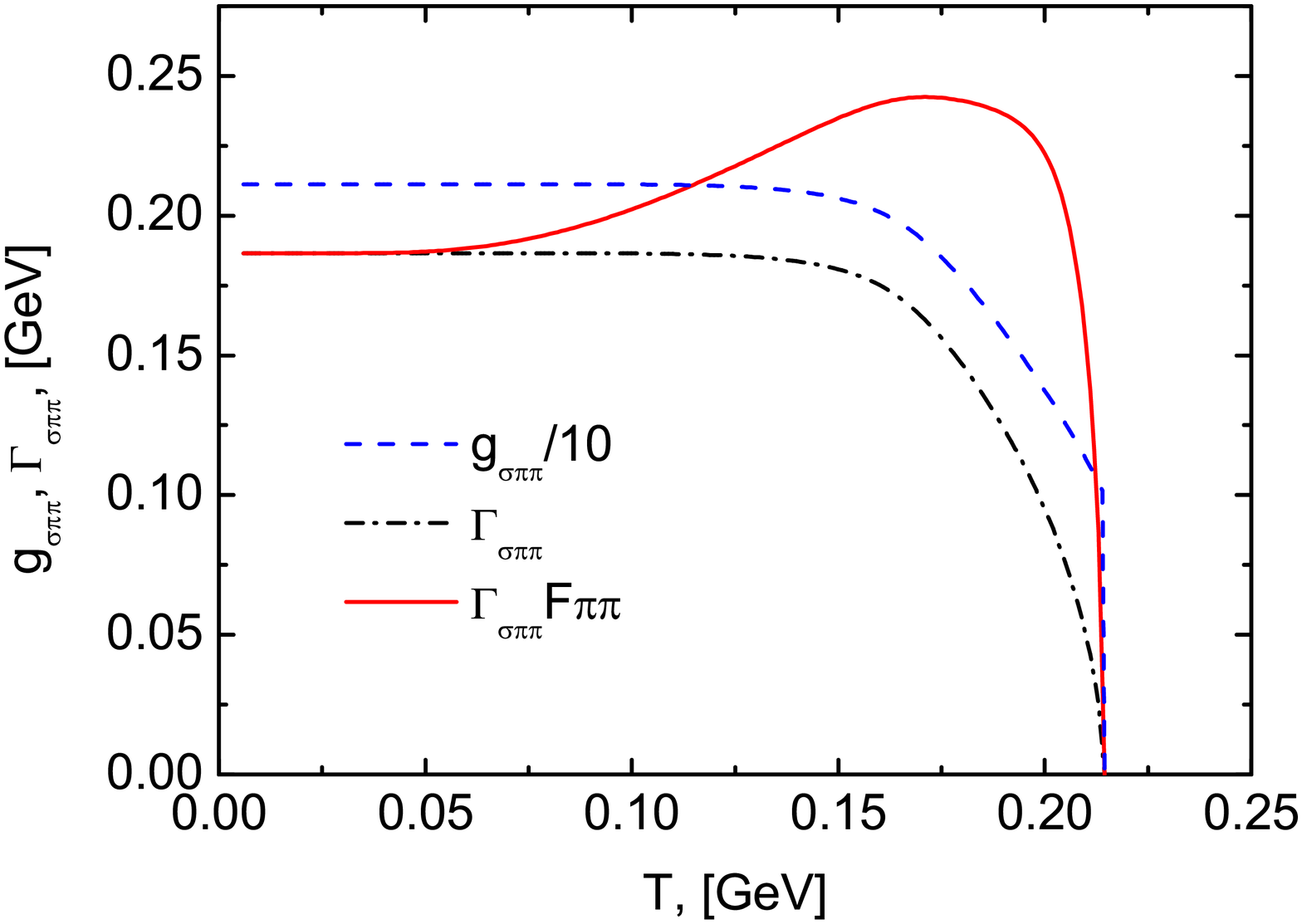}}
\caption{The temperature dependence of the total decay width
$\Gamma_{\sigma\rightarrow\pi\pi}$
and the coupling strength $g_{\sigma\pi\pi}$  at the zero chemical potential. 
The solid and dash-dotted lines are the results for the decay width
(\ref{Gam0}) with and without the Bose-Einstein factor $F_{\pi\pi}$, respectively.}
 \label{fdec}
\end{figure}

\begin{figure}[h!]
\centerline{
\includegraphics[width = 7. cm ]{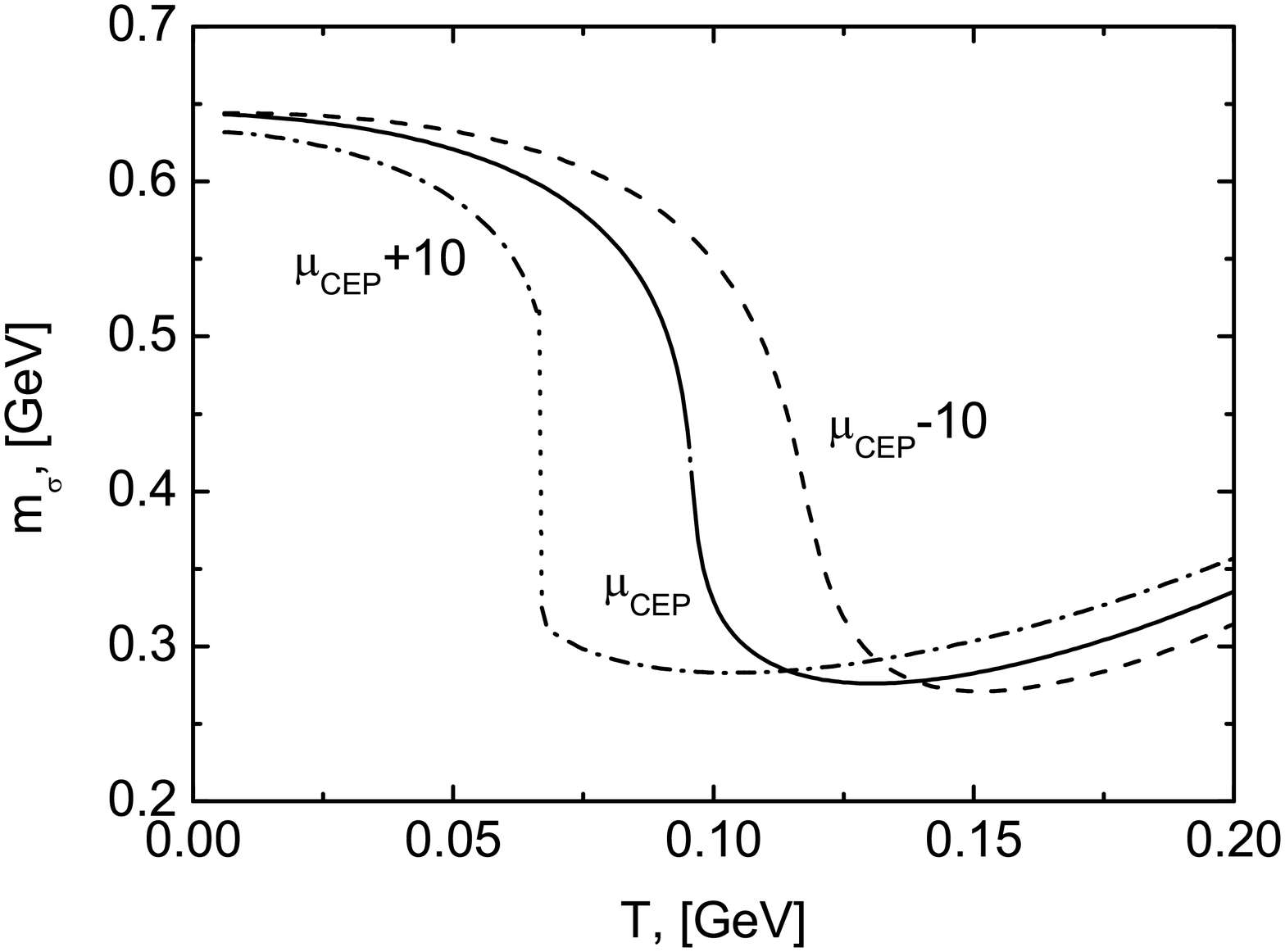} \hspace*{-1.0cm}
\includegraphics[width = 7. cm ]{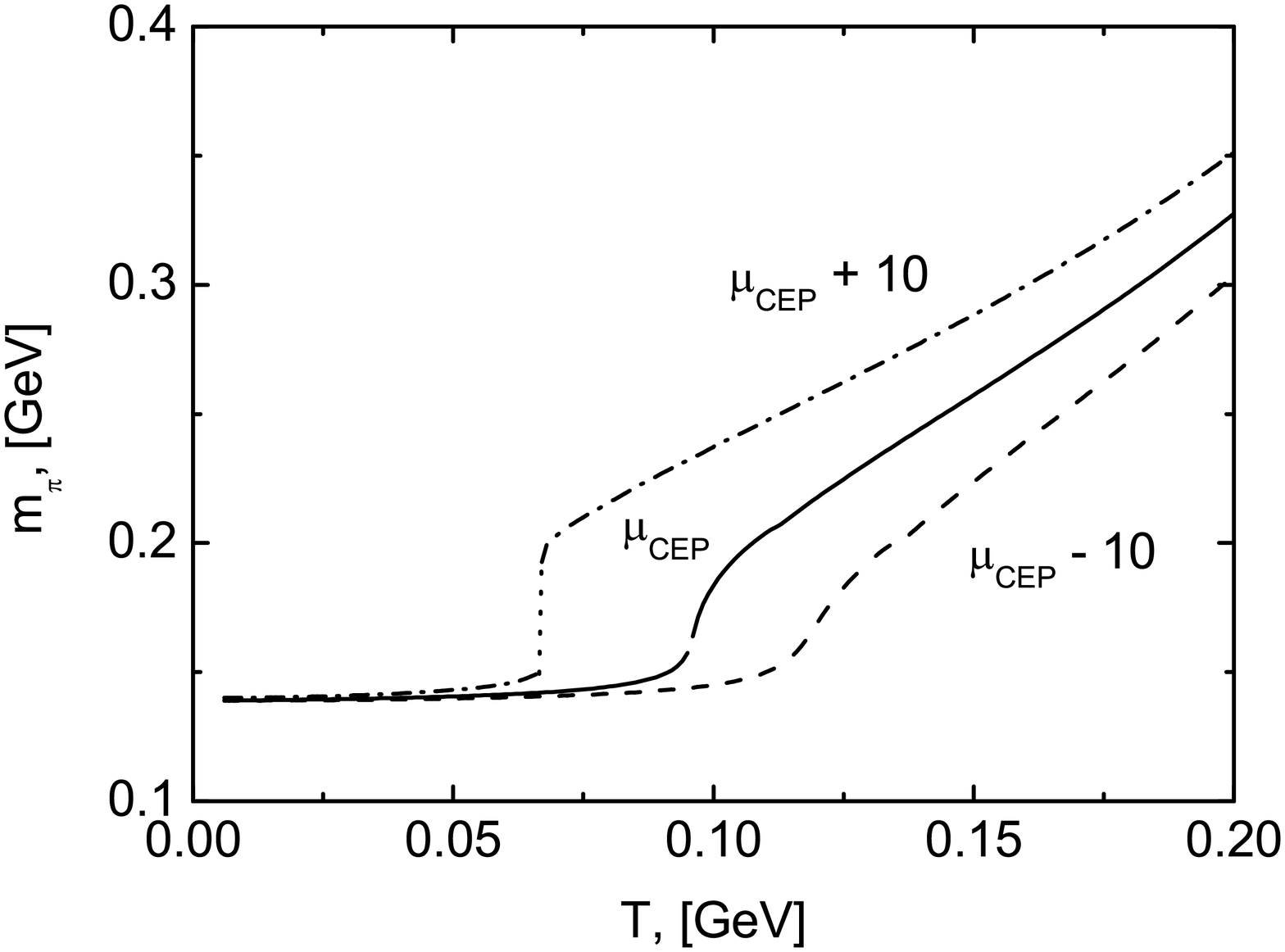} }
\caption{Temperature dependence of $\sigma$ (left panel) and $\pi$ (right panel)
meson masses near the critical end-point. Solid lines correspond to the
chemical potential at the critical end-point $\mu_{CEP}$. Dot-dashed and dashed
lines are the PNJL results for $\mu_{CEP}+10$ and $\mu_{CEP}-10$ MeV, respectively.
The dotted lines correspond to the mixed phase.
}
 \label{massB}
\end{figure}

\begin{figure}[h!]
\centerline{
\includegraphics[width = 7. cm ]{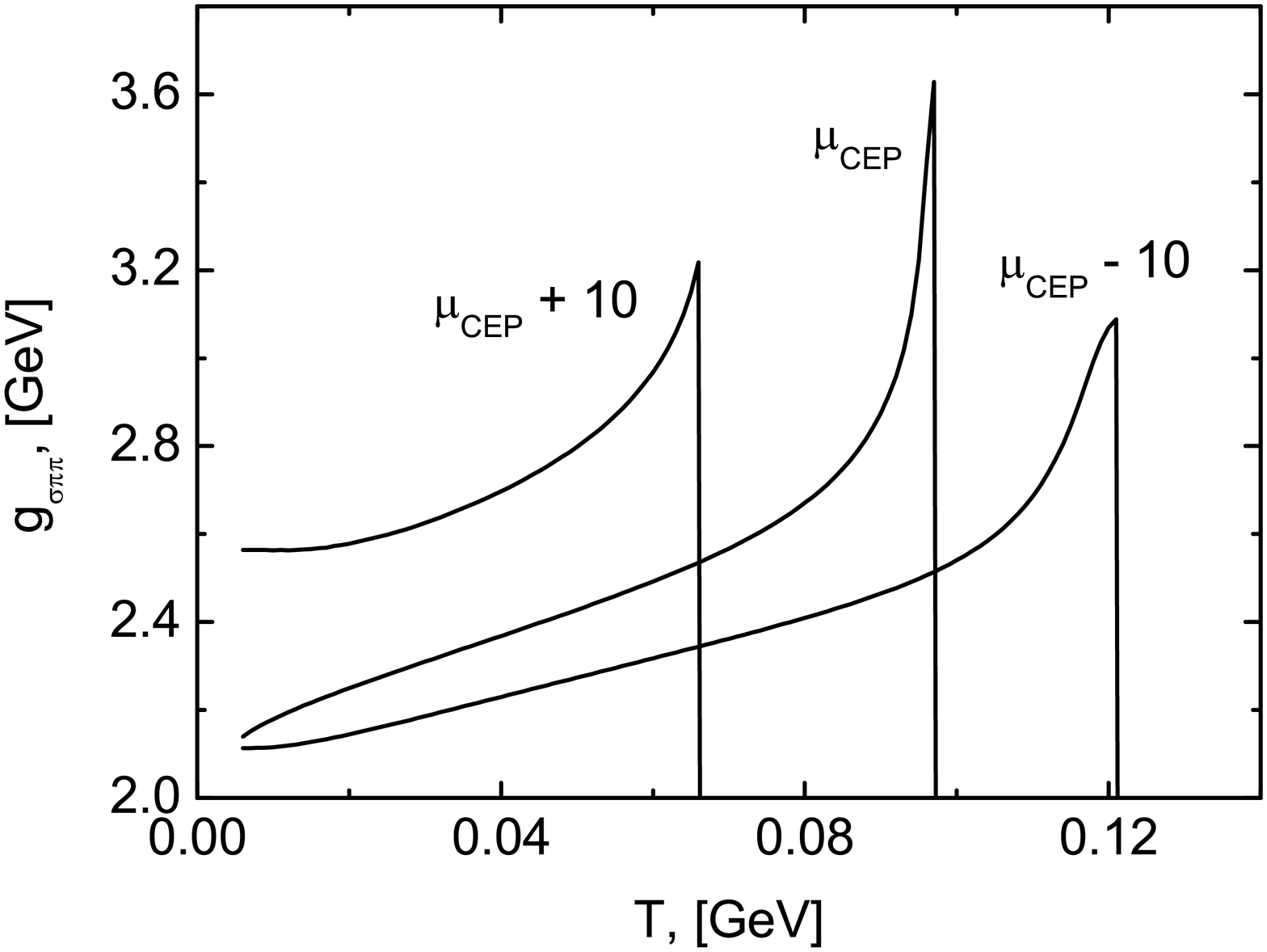}\hspace*{-1.0cm}
\includegraphics[width = 7. cm ]{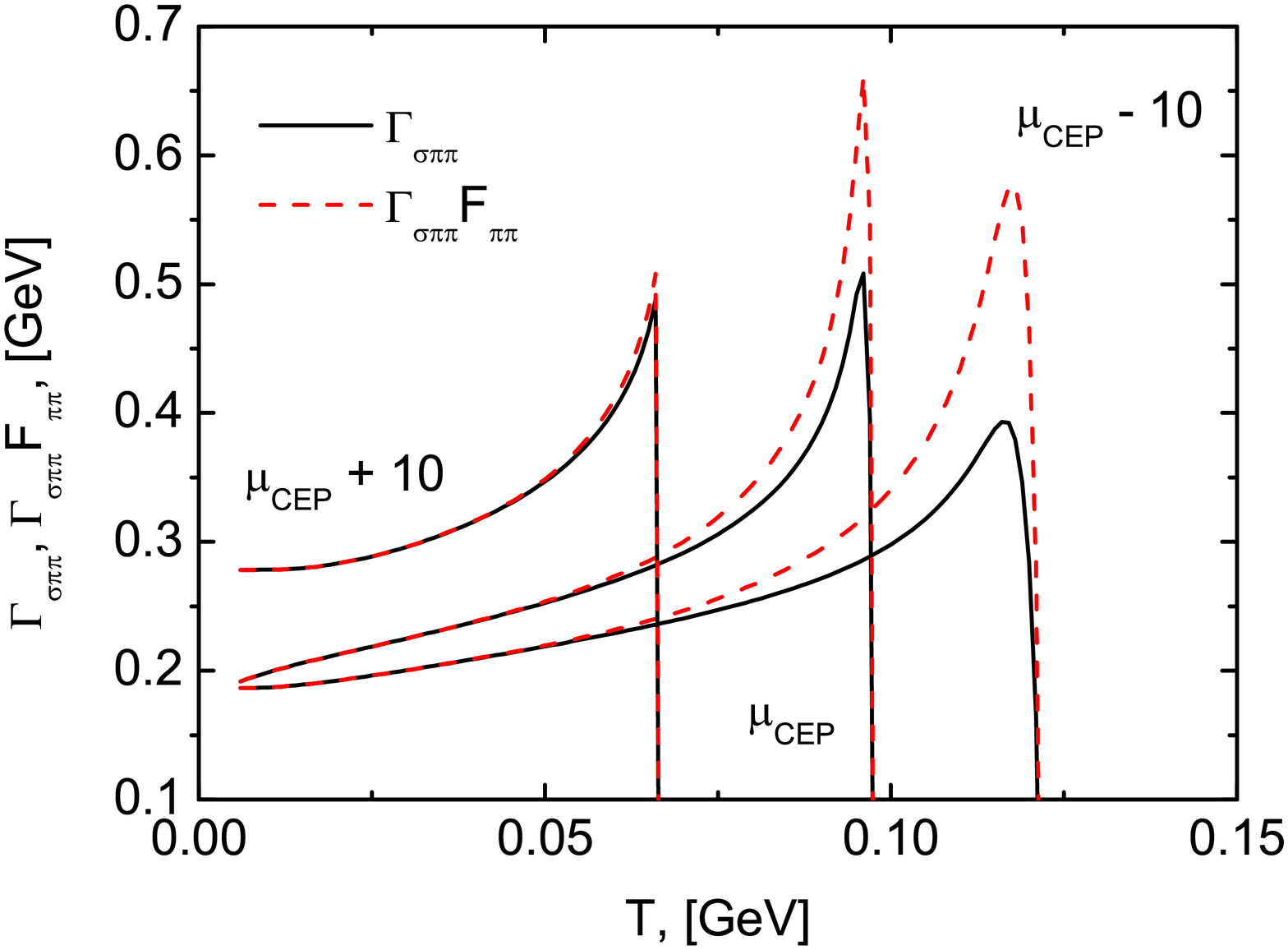}}
\caption{Left panel: The temperature dependence of
the coupling strength $g_{\sigma\pi\pi}$  at three values of the chemical potential
$\mu = \mu_{CEP}$, $\mu_{CEP}$-10 MeV and $\mu_{CEP}$+10 MeV.
Right panel: The total decay width $\Gamma_{\sigma\rightarrow\pi\pi}$ at the same
values of chemical potentials. The total decay width
$\Gamma_{\sigma\pi\pi} \ F_{\pi\pi}$ and  the width $\Gamma_{\sigma\pi\pi}$,
Eq.(\ref{Gam0}) are plotted by the solid  and dashed lines, respectively,
for the same three values of $\mu$.}
 \label{deccep}
\end{figure}

 For the  existing decay
$\sigma\rightarrow\pi\pi$ the kinematic factor
$\sqrt{1-{4m_{\pi}^2}/{m_{\sigma}^2}}$ in
(\ref{Gam0})  leads to the constraint  $m_{\sigma} \leq
2m_{\pi}$ which is $T$- and $\mu$-dependent.  One may expect that in
the temperature region where
this condition is broken the values $g_{\sigma\pi\pi}$
and $\Gamma_{\sigma\rightarrow\pi\pi}$ will drop to zero (Fig.~\ref{fdec}).
In the  case $\mu=0$ considered the kinematic condition is broken at the 
$\sigma\to\pi\pi$ dissociation temperature $T_d^\sigma\approx$ 210 MeV.

The coupling constant $g_{\sigma \pi \pi}$ is about $2.1$ GeV (note
the scaling factor 1/10 in Fig.\ref{fdec}) in vacuum and  stays almost
constant up to $T\leqslant 0.22$ GeV (at $\mu = 0$) and then
it drops to zero at  $m_{\sigma} = 2m_{\pi}$. The experimental value, extracted
from the J/$\psi$ decays, $g_{\sigma\pi\pi}=2.0^{+0.3}_{-0.9}$ GeV~\cite{BES01}
is in reasonable agreement with our result. It is of interest to note that the  
quark-meson models predict $g_{\sigma\pi\pi}=$1.8 GeV~\cite{FGILW03} and
$1.8^{+0.5}_{-0.3}$ GeV~\cite{HZH02},  and the linear sigma model gives 
 $g_{\sigma\pi\pi}=2.54\pm$ 0.01 GeV~\cite{DR01}.

The total $\sigma$ decay width was measured recently in two
experiments~\cite{BES01,HZH02,E791}.
The Beijing Spectrometer (BES) Collaboration at the
Beijing Electron-Positron Collider reported evidence of the existence of the
$\sigma$ particle in J/$\psi$ decays. In the $\pi^+\pi^-$ invariant mass spectrum
in the process of the J/$\psi \to \sigma\omega\to \pi^+\pi^-\omega$
 they found a low mass enhancement, and the
detailed analysis strongly favors $O^{++}$ spin parity with a statistical
significance for the existence of the $\sigma$ particle.  The BES measured
values of the $\sigma$  mass and total width are~\cite{BES01,HZH02}
\begin{eqnarray} \label{exp1}
m_\sigma=390^{+60}_{-36} \ MeV, \hspace*{1cm}
\Gamma_{\sigma\to\pi\pi}= 282_{-50}^{+77} \ MeV.
\end{eqnarray}
The E791 Collaboration at Fermilab reported on evidence of a light and broad
scalar resonance in the nonleptonic cascade decays of heavy mesons~\cite{FGILW03}.
It was found in the Fermilab experiment that the $\sigma$ meson is rather
important in the $D$ meson decay $D\to3\pi$ generated by the intermediate
$\sigma$-resonance channel
\begin{eqnarray} \label{exp2}
m_\sigma=478^{+24}_{-23}\pm17 \ MeV, \hspace*{1cm}
\Gamma_{\sigma\to\pi\pi}= 324_{-42}^{+40}\pm 21 \ MeV.
\end{eqnarray}

These experimental values should be compared with
$\Gamma_{\sigma\to\pi\pi}\simeq$190 MeV at $T=\mu=$0 (see Fig.\ref{fdec}).
One can additionally  take into account the Bose-Einstein statistics of
 final pion states by introducing the factor   $F_{\pi\pi} =
(1+f_B(\frac{m_\sigma}{2}))^2$~\cite{zhuang}  into the total decay width (\ref{Gam0}),
where the boson distribution function $f_B(x) = (e^{x/T} - 1)^{-1}$. In contrast to
the kinematic factor $\sqrt{1-{4m_{\pi}^2}/{m_{\sigma}^2}}$,
this pion distribution function tends to increase the width. Near the Mott
temperature $\Gamma_{\sigma\to\pi\pi}\approx$250 MeV at $\mu=$0.
In fact, the numerical calculation shows that $\Gamma_{\sigma\pi\pi}(T)$ decreases
as $T$ goes up, and eventually vanishes at a high temperature. The measured widths
(\ref{exp1}),(\ref{exp2}) are somewhat higher then that in our model at $T=$0.
It is noteworthy that the measured $\sigma$ masses $390^{+60}_{-36}$~\cite{BES01}
and $478^{+24}_{-23}$~\cite{E791} are noticeably smaller than in our model
$m_\sigma\approx$620  MeV  (to be fixed by the model parameters) while the total decay
width  depends strongly on $m_\sigma$, see Eq.(\ref{Gam0}).  The quark-meson models
give the decay width that does not differ essentially from our estimate:
$\Gamma_{\sigma\to\pi\pi}=$173~\cite{FGILW03} and 149.9~\cite{MA10} MeV though
the used sigma meson mass is close to experimental ones, being $m_\sigma=$
485.5 and 478 MeV, respectively.

The decay width was really not studied at the nonzero baryon density, 
in particular, in the region near the critical end-point. For the finite
$\mu$ the coupling strength and meson masses behave in a nontrivial way.
Both  $\sigma$ and $\pi$ mesons suffer a jump in the region of the first-order
phase transition (see curve for $\mu+10$ MeV in Fig.\ref{massB}) which ends
at the critical end-point $\mu_{CEP}$ and then they change continuously for
$\mu<\mu_{CEP}$, where the crossover-type phase transition occurs.

According to (\ref{Gam0}), this mass behavior defines the total
decay rate of a $\sigma$ meson at $\mu \neq 0$, as shown in the
right panel of Fig.~\ref{deccep}. As can be seen, the $T$-region
of the decay enhancement becomes narrower with increasing $\mu$.
The total widths $\Gamma_{\sigma\to\pi\pi}$ and
$\Gamma_{\sigma\to\pi\pi} F_{\pi\pi}$ tend to grow with
temperature exhibiting a narrow maximum due to particularities of
the $g_{\sigma\pi\pi}^2$ term in~(\ref{Gam0}). The inclusion of the
$F_{\pi\pi}$ factor enhances this maximum, but  this effect becomes
smaller when one moves by $\Delta \mu$ toward larger $\mu$ from 
$\mu_{CEP}$ because the pion mass increases above $\mu_{CEP}$ (see
Fig.\ref{massB}). If the chemical potential $\mu=\mu_{CEP}$, the
temperature at which coupling strength $g_{\sigma\pi\pi}$ or total
decay width $\Gamma_{\sigma\to\pi\pi}$ drops down to zero, the
temperature is just $T=T_{CEP}$. If $\mu$ increases (decreases)
with respect to $\mu_{CEP}$ by $\Delta \mu\sim$10 MeV, the
temperature decreases (increases), respectively, by about 30 MeV
from $T_{CEP}$. Maximal values of $\Gamma_{\sigma\to\pi\pi}$ near the 
critical end-point are larger than that for $\mu=$0 by a factor about 3. 
It means that a $\sigma$ meson lives shorter in the dense baryon matter. 
 The shape of $g_{\sigma\pi\pi}(T)$ is insensitive
to $\mu$ around $\mu_{CEP}$. The full decay width
$\Gamma_{\sigma\to\pi\pi}$ exhibits some wide maximum near the
temperature $T_{CEP}$. This maximum is enhanced due to the $F_{\pi\pi}$
factor and gets more pronounced for smaller $\mu$.

\section*{Concluding remarks}

The two-flavor PNJL model that reasonably describes quark-meson
thermodynamics at finite temperature and chemical potential is
applied to calculate the $\sigma\to \pi\pi$ decay in this
medium. The emphasis here is made on the behavior near the
critical end-point. At $\mu=$0 and $T\to$0 the coupling constant
$g_{\sigma\pi\pi}=$2.1 GeV and the total $\sigma$ decay width
$\Gamma_{\sigma\to\pi\pi}\approx$190 MeV are in reasonable agreement
with both available experimental data and quark-meson model
estimates. At finite quark chemical potential near the critical
end-point the $\Gamma_{\sigma\to\pi\pi}$ width shows a sharp
maximum coming from a particular behavior  of the coupling strength
$g_{\sigma\pi\pi}$. The sigma mesons live here a shorter time than in
the baryonless matter. The rapid decrease of
$\Gamma_{\sigma\to\pi\pi}$ at a high temperature is due to the phase
space factor (see Eq.(\ref{Gam0})). The account for the
Bose-Einstein statistics in the final state pions (the factor
$F_{\pi\pi}$) results in the appearance of some  nonthermal maximum of
the decay width near $T$ and $\mu$ at which the kinematic
condition $m_{\sigma} \leq 2m_{\pi}$ is broken. This width
enhancement is about  $\sim 20\%$ at $\mu_{CEP}$ and is negligible
if one moves to $\mu=\mu_{CEP}+\Delta \mu$.

The presented results are obtained in the first order in the
$1/N_c$ expansion. In a more realistic case the $\sigma-\omega$
and $\sigma-A_1$ mixing  can affect noticeably the considered
quantities, especially for the $\mu\ne $0 case~\cite{STT98}. 
However, it corresponds to higher orders in $1/N_c$.

The measurements of nonthermal enhancement of pions might be
considered as a signature of chiral phase transition. However, it
is a difficult experimental problem since the $\sigma$ life time is very 
short, and the pion contribution from the
resonance decay should be separated carefully. So a more
elaborated analysis is needed.

\section*{Acknowledgments}

We are grateful to P. Costa,  E.A. Kuraev, V.V. Skokov, and M.K.
Volkov for useful comments. V.T acknowledges the financial support
within the ``HIC for FAIR'' center of the ``LOEWE'' program and
Heisenberg-Landau grant. The work of Yu. K. was supported by the
RFFI grant 09-01-00770a.

\end{document}